\newcommand{\myname}{Geoff Boeing}
\newcommand{\myemail}{boeing@usc.edu}
\newcommand{\myaffiliation}{Department of Urban Planning and Spatial Analysis\\University of Southern California}
\newcommand{\paperdate}{March 2021}
\newcommand{\papertitle}{Street Network Models and Indicators for Every Urban Area in the World}
\newcommand{\papercitation}{Boeing, G. 2021. \papertitle. \emph{Geographical Analysis}, published online ahead of print. \href{https://doi.org/10.1111/gean.12281}{doi:10.1111/gean.12281}}
\newcommand{\paperkeywords}{Urban Planning, Transportation, Data Science, Street Networks}

\RequirePackage[l2tabu,orthodox]{nag}   
\documentclass[11pt,letterpaper]{article} 

\usepackage[T1]{fontenc}                
\usepackage[utf8]{inputenc}             
\usepackage{crimson}                    
\usepackage{helvet}                     

\usepackage[strict,autostyle]{csquotes} 
\usepackage[USenglish]{babel}           
\usepackage{microtype}                  

\usepackage{abstract}                   
\usepackage{authblk}                    
\usepackage{booktabs}                   
\usepackage{caption}                    
\usepackage[final]{draftwatermark}      
\usepackage{endnotes}                   
\usepackage{geometry}                   
\usepackage{graphicx}                   
\usepackage{hyperref}                   
\usepackage{natbib}                     
\usepackage{pdflscape}                  
\usepackage{setspace}                   
\usepackage{titlesec}                   
\usepackage{url}                        

\graphicspath{{./figures/}}

\titleformat{\section}{\normalfont\sffamily\large\bfseries\color{black}}{\thesection.}{0.3em}{}
\titleformat{\subsection}{\normalfont\sffamily\small\bfseries\color{black}}{\thesubsection.}{0.3em}{}

\hypersetup{
	pdfauthor={\myname},
	pdftitle={\papertitle},
	pdfsubject={\papertitle},
	pdfkeywords={\paperkeywords},
	pdffitwindow=true,         
	breaklinks=true,           
	colorlinks=false,          
	pdfborder={0 0 0}          
}

\SetWatermarkText{DRAFT}
\SetWatermarkScale{1.3}
\SetWatermarkLightness{0.9}

\begin{document}

	\title{\papertitle\thanks{{Preprint of: \papercitation}}}
	\author{\myname}
	\affil{\myaffiliation}
	\date{\paperdate}

	\maketitle

\begin{abstract}

Cities worldwide exhibit a variety of street network patterns and configurations that shape human mobility, equity, health, and livelihoods. This study models and analyzes the street networks of every urban area in the world, using boundaries derived from the Global Human Settlement Layer. Street network data are acquired and modeled from OpenStreetMap with the open-source OSMnx software. In total, this study models over 160 million OpenStreetMap street network nodes and over 320 million edges across 8,914 urban areas in 178 countries, and attaches elevation and grade data. This article presents the study's reproducible computational workflow, introduces two new open data repositories of ready-to-use global street network models and calculated indicators, and discusses summary findings on street network form worldwide. It makes four contributions. First, it reports the methodological advances of this open-source workflow. Second, it produces an open data repository containing street network models for each urban area. Third, it analyzes these models to produce an open data repository containing street network form indicators for each urban area. No such global urban street network indicator dataset has previously existed. Fourth, it presents a summary analysis of urban street network form, reporting the first such worldwide results in the literature.

\end{abstract}

\section{Introduction}

Street networks shape the city. They structure the circulation patterns of people and goods and underlie urban accessibility. Differences in street network geometry and topology---collectively, \enquote{form}---worldwide reflect different cultures, political systems, urbanization eras, technology, design paradigms, climates, and terrain. These networks in turn organize physical urban space and influence the ability to traverse it via different modes of transportation.

Yet in this era of post-globalization, little is known about comparative street network form worldwide at the urban scale. This is largely due to data access and computational limitations. Traditionally, delineating consistent urban area boundaries was difficult, making it challenging to define consistent study sites \citep{lemoine-rodriguez_global_2020}. Even if consistent study sites could be established, it was nearly impossible to gather consistent comprehensive street network data around the world. And even if one did, it was nearly impossible to manipulate and organize the hundreds of millions of geospatial elements that that would entail, then model them in a graph-theoretic way, then compute geometric and topological indicators of form. Nevertheless, such models, indicators, and analyses would be useful for understanding urbanization patterns, transportation infrastructure planning, and the path to sustainable urban form for cities worldwide.

This study takes advantage of several emerging tools, technologies, and open data to model the individual street networks of every urban area in the world, compute geometric and topological indicators, and analyze them. It uses the Global Human Settlement Layer (GHSL) to define urban area boundaries and other variables. Using OSMnx, it downloads and models urban-scale street network data globally from OpenStreetMap, attaches elevation data, then calculates indicators for each urban area. It places all of the resulting network models, indicators, and code into open data repositories for public reuse.

This article reports the novel methods employed in this modeling and analytics project. Then it documents the street network model repository and its contents and the indicators repository and its contents. Next it presents a high-level analysis of worldwide urban street network form, using these models and indicators. Across all urban areas worldwide, both total street length and intersection count scale sublinearly with population. Higher per capita GDP is associated with higher per capita total street length. Validation reveals that the open data used as elevation attributes compare favorably with high-quality commercial data. The article concludes with notes on reuse. In sum, this study produces the first comprehensive public data repository of ready-to-use urban street network models and indicators worldwide and reports the first such worldwide analytical results.

\section{Background}

\subsection{Street Network Models}

Street network models come in many flavors, but most commonly are mathematical models called graphs \citep{newman_structure_2003,newman_networks:_2010,trudeau_introduction_1994,vespignani_twenty_2018,brandes_network_2005,gastner_spatial_2006}. Graphs can represent both the geometry and the topology of a real-world street network. Abstractly, a graph $G$ comprises a set of nodes (i.e., elements) $N$ which are linked to one another by a set of edges (i.e., connections) $E$. Each edge $e$ in set $E$ either connects two nodes or connects a single node to itself as a self-loop. Parallel edges exist when multiple edges connect the same two nodes.

Network modelers must decide on several theoretical aspects of representation, including directedness, planarity, and primality \citep{fischer_spatial_2014,marshall_street_2018}. In the case of a directed graph, all the edges in $E$ point one-way from some node $u$ to another node $v$. This may allow for the possibility of a self-loop where $u=v$. In the case of an undirected graph, all the edges in $E$ point bidirectionally between the nodes they link. If a graph is planar, all the edges in $E$ intersect in a two-dimensional plane exclusively at nodes in $N$. If this condition does not hold, the graph is nonplanar \citep{cardillo_structural_2006,hopcroft_efficient_1974,masucci_random_2009,szekely_successful_2004,viana_simplicity_2013}. A primal graph of a street network models intersections and dead-ends as nodes and the street segments that connect them as edges \citep{porta_network_2006-1}. A dual graph of a street network does the opposite, modeling street segments as nodes and intersections as edges \citep{porta_network_2006}. Real-world street networks often have self-loops, parallel edges, flow directionality restrictions such as one-way streets, and nonplanar elements such as overpasses and underpasses \citep{boeing_osmnx:_2017}.

\subsection{Street Network Data}

Data on street networks around the world exist in various sources of various quality and accessibility. Many are digitized by local or regional authorities, resulting in inconsistencies in digitization standards, spatial validity, attribute data quality, and file formatting. High-quality street network geometry data exist for most developed countries, but data inconsistencies and language barriers make international cross-sectional comparison difficult. Furthermore, most such datasets exist in shapefile format and thus contain network geometry but minimal information about topology. Yet both geometry and topology are essential to consider in most spatial network analyses. Street networks are spatially embedded and are thus defined by both their geometry (e.g., positions, lengths, areas, angles, etc.) and their topology (i.e., connections and configurations) \citep{barthelemy_spatial_2011,fischer_spatial_2014}.

Given these limitations, better data sources and network models are important for international street network analysis. Four key areas of improvement would include: 1) global coverage and availability, 2) consistent digitization and attribute data, 3) consistent representation of both geometric and topological data, and 4) better public accessibility and usability. Online geographic information systems, volunteered geographic information, and crowd-sourced big data create new opportunities to address these points. In particular, OpenStreetMap offers an important alternative source of street network data \citep{jokar_arsanjani_openstreetmap_2015}.

OpenStreetMap is an open-source, collaborative, worldwide mapping project and database. One can query its database for street and intersection data, along with attribute data about road types, names, and (when available) speeds, widths, and numbers of lanes. It offers good global coverage and high geometric and topological data quality \citep{girres_quality_2010,haklay_how_2010,corcoran_analysing_2013,zielstra_assessing_2013,barron_comprehensive_2014,maier_openstreetmap_2014,basiri_quality_2016,sehra_extending_2020}. \citet{barrington-leigh_worlds_2017} found that, as of 2016, OpenStreetMap was 83\% complete worldwide, over 40\% of countries' (including many developing countries) street networks were effectively 100\% complete, and completeness was highest in both dense cities and sparsely populated areas.

As of 2021, OpenStreetMap has more than 7 million contributors who have added over 6.6 billion nodes (points), 730 million ways (lines and boundaries), and attendant attribute data to its database. Volunteers provide editorial oversight of contributions and changes. However, despite its large user base, researchers estimate that >95\% of these contributors are male, and as such, there may be correlated biases in the contributed content \citep{schmidt_gender_2013,graham_towards_2015}. While OpenStreetMap road coverage is generally good worldwide, other geospatial features have better coverage in developed countries and in cities versus small towns. No data source is perfect, but OpenStreetMap is global, publicly-accessible, free, and an Open Source Initiative affiliate.

Accordingly, OpenStreetMap helps address the first three of the four areas of improvement listed earlier. However, the fourth problem persists: it is not particularly accessible or usable for less-technical urban scholars to use its data for graph models and analytics. Researchers usually acquire OpenStreetMap data through its APIs or by downloading a prepackaged data extract from a third-party. Either option offers useful raw data but usually requires writing and testing hundreds of lines of code to process topological relations and construct graph models. Doing this on an ad hoc basis introduces challenges for interpretation and replication as many small computational and modeling decisions get made along the way, such as the exact handling of common street network features like self-loops, parallel edges, and culs-de-sac. A consistent set of well-documented models and indicators, generated with an accessible open-source workflow, would improve what is otherwise often a black box \citep{boeing_right_2020}.

\subsection{Street Network Indicators}

Several efforts in recent years have aimed to address these challenges and generate sets of urban street network indicators. For example, the OSMnx project takes this motivation to develop an open-source Python package for automatically downloading, modeling, and analyzing street networks and other geospatial features from OpenStreetMap \citep{boeing_osmnx:_2017}. Using this tool, a recent project modeled the street networks of every US city/town, county, urbanized area, census tract, and Zillow-defined neighborhood, placed these models online in a public open data repository, and conducted spatial network analyses on them \citep{boeing_multi-scale_2020}.

Similarly, \citet{dingil_transport_2018} used OSMnx to calculate transportation indicators for 151 urban areas worldwide. \citet{karduni_protocol_2016} created a data repository with 80 worldwide cities' street networks derived from OpenStreetMap data. \citet{da_cruz_metropolitan_2020} developed a database of urban indicators across 58 metropolitan areas worldwide, but did not include street network form indicators. \citet{barrington-leigh_global_2019,barrington-leigh_global_2020} used all the streets mapped in OpenStreetMap to generate global indicators of street network disconnectivity to explore cross-sectional and longitudinal trends in urban sprawl.

\section{Methods}

\subsection{Urban Area Boundaries}

The present study builds on this past work to model and analyze the street networks of every urban area in the world. It defines these units of analysis using spatial boundaries derived from the publicly available GHSL Urban Centre Database\endnote{UCD: \url{http://data.europa.eu/89h/53473144-b88c-44bc-b4a3-4583ed1f547e}} (UCD) version 2019a 1.2, a project supported by the European Commission's Joint Research Centre and Directorate-General for Regional and Urban Policy \citep{florczyk_description_2019}. In addition to these urban area boundaries, the UCD provides attribute data such as the names of the country and core city, population, built-up area, gross domestic product (GDP), UN income class and development group, transport-sector emissions, particulate matter concentration, climate, and land use efficiency.

The GHSL project uses spatial data mining to organize a vast amount of data from satellite image streams, censuses, and volunteered geographic information. Its UCD data product delineates urban areas (which it calls \enquote{urban centres}) using these data from the GHSL and other scientific open data sources. It defines these urban areas using resident population and built-up surface across a global 1 km\textsuperscript{2} grid, using the DEGURBA method of delineating urban/rural areas for international statistical comparison, developed jointly by the European Commission, the World Bank, the Organization for Economic Cooperation and Development, the UN Food and Agriculture Organization, and the UN Human Settlements Programme. Thus the UCD consists of \enquote{high-density clusters of contiguous grid cells of 1 km\textsuperscript{2} with a density of at least 1,500 inhabitants per km\textsuperscript{2} and a minimum population of 50,000} \citep[][p. 13]{florczyk_description_2019}.

\subsection{Graph Modeling}

This study uses OSMnx\endnote{The data were downloaded with OSMnx v1.0.0 in January 2021} to download street network data from OpenStreetMap and construct graph models of each urban area's drivable street network. It does this for every urban area in the UCD that satisfies three conditions: 1) is marked \enquote{true positive} in the UCD, 2) has $\ge1$ km\textsuperscript{2} built-up area, and 3) includes at least three OpenStreetMap drivable street network nodes within its boundaries. This comprises 8,914 total urban areas.

This study models these street networks as nonplanar directed multigraphs with possible self-loops. All of these models are primal graphs to account for the full geographic characteristics of the street network \citep{ratti_space_2004,batty_network_2005}. The workflow retains all graph components even if they are not fully connected and is parameterized to retrieve all public drivable streets, excluding service roads like alleyways or parking lot circulation. This parameterization includes living streets, shared streets, woonerfs, and the like in the models, but does exclude streets and other pathways where motor traffic is forbidden, which may impact some cities more than others. Additionally, spatial graphs often exhibit adverse periphery effects due to an artificial boundary being imposed: OSMnx attenuates some of these by initially downloading and modeling a larger area than requested to correctly calculate node degrees before removing peripheral nodes and edges that fall outside the requested boundary polygon.

\begin{figure*}[btp]
	\centering
	\includegraphics[width=0.7\textwidth]{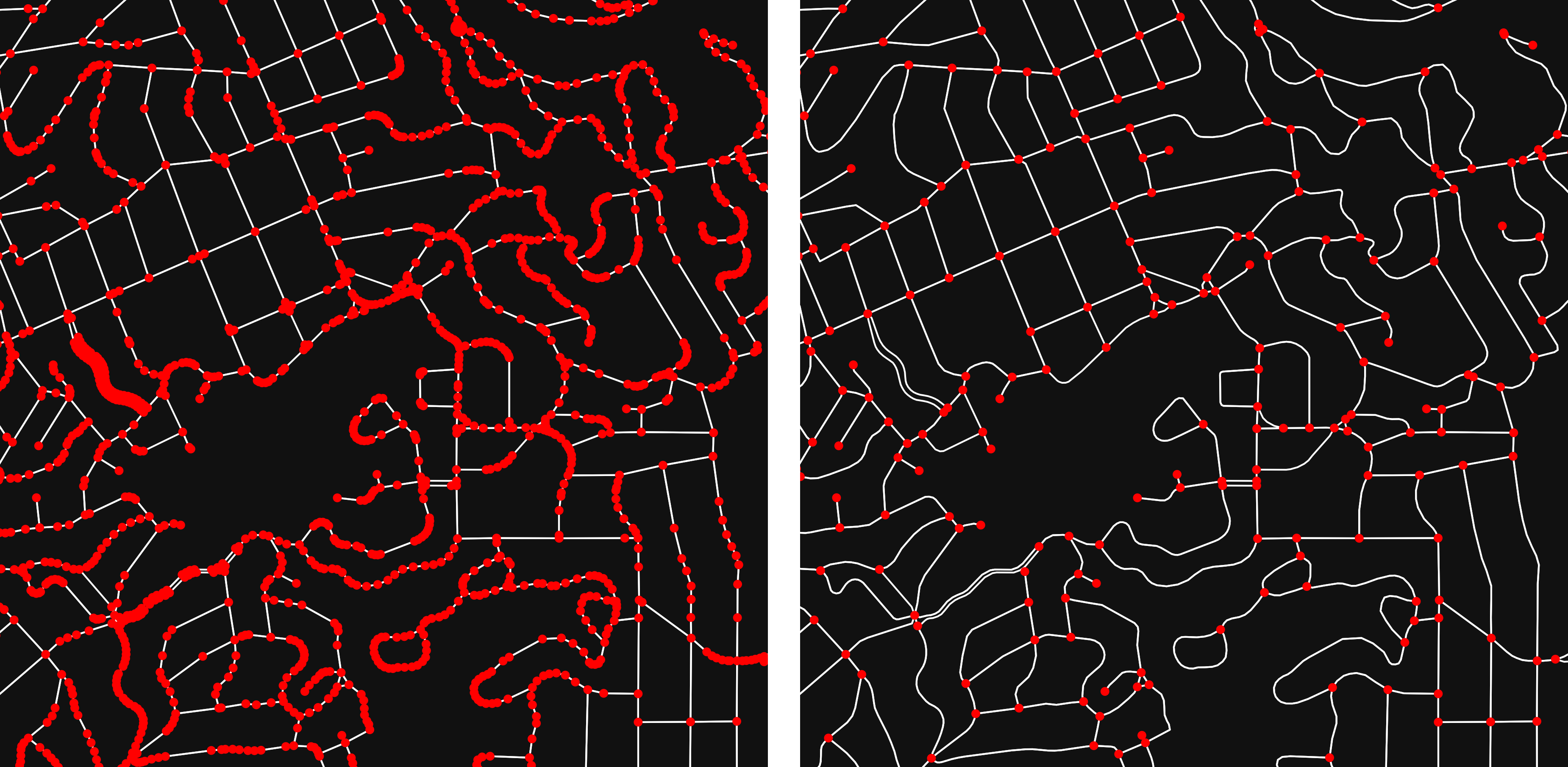}
	\caption{The graph of a town's street network before (left) and after (right) topological simplification. Circles are nodes and lines are edges.}
	\label{fig:simplification}
\end{figure*}

This yields a set of models collectively comprising over 160 million nodes and 320 million edges. For better theoretical correspondence, OSMnx next topologically simplifies the graphs to retain nodes only at true intersections and dead-ends, while retaining the true spatial geometry of each edge (i.e., street segment) between them \citep{boeing_osmnx:_2017}. This is a crucial step before conducting analytics with OpenStreetMap network data, such as calculating intersection density or average node degree. Raw OpenStreetMap data represent nodes as geometric vertices of straight-line segments composing more complex lines. Simplification produces a model that corresponds better to graph theory and transportation geography with nodes representing intersections and dead-ends and edges representing street segments. See Figure \ref{fig:simplification}. Simplification yields a final set of models collectively comprising 37 million nodes and 53 million edges.

\subsection{Elevation}

Next, we attach elevation above sea level to every node in every graph. These elevations come from two publicly available digital elevation models (DEMs): the Advanced Spaceborne Thermal Emission and Reflection Radiometer (ASTER) v2 and the Shuttle Radar Topography Mission (SRTM). The ASTER DEM covers the world from 83° N to 83° S at a spatial resolution of one arcsecond (roughly 30 meters at the equator). The SRTM DEM covers the world from 60° N to 56° S at a spatial resolution of three arcseconds (roughly 90 meters at the equator). The ASTER DEM is finer resolution but exhibits more noise. The SRTM DEM is coarser resolution but less noisy, and was further processed by CGIAR-CSI to fix errors and fill voids \citep{gorokhovich_accuracy_2006}. We additionally collect the elevation of each node using the Google Maps Elevation API, but only for validation purposes as Google provides global and high-quality but commercial and closed-source data with restrictive licensing \citep[cf.][]{rusli_google_2014}.

On a node-by-node basis, the workflow selects either the ASTER or SRTM value to assign as the node's elevation. It selects for each node whichever of the ASTER or SRTM elevation values has the least absolute difference from the corresponding Google value, as a validation reference point. Finally it calculates the grade (i.e., incline) of each edge then saves each urban area's graph model to disk as a GeoPackage file, a GraphML file, and node/edge lists in comma-separated values (CSV) format.

\begin{table}[htbp]
	\centering
	\footnotesize
	\caption{Fields in the indicators dataset. Those carried over from the UCD are noted as such.}
	\label{tab:indicators}
	\begin{tabular}{p{3.8cm} p{1.6cm} p{6.6cm}}
		\toprule
		Indicator Name                & Type    & Description \\
		\midrule
		country                       & string  & Primary country name \\
		country\_iso                  & string  & Primary country ISO 3166-1 alpha-3 code \\
		core\_city                    & string  & Urban area core city name \\
		uc\_id                        & integer & Urban area's unique identifier in UCD \\
		cc\_avg\_dir                  & decimal & Avg clustering coefficient (unweighted, directed) \\
		cc\_avg\_undir                & decimal & Avg clustering coefficient (unweighted, undirected) \\
		cc\_wt\_avg\_dir              & decimal & Avg clustering coefficient (weighted, directed) \\
		cc\_wt\_avg\_undir            & decimal & Avg clustering coefficient (weighted, undirected) \\
		circuity                      & decimal & Ratio of street lengths to straight-line distances \\
		elev\_iqr                     & decimal & Interquartile range of node elevations, meters \\
		elev\_mean                    & decimal & Mean node elevation, meters \\
		elev\_median                  & decimal & Median node elevation, meters \\
		elev\_range                   & decimal & Range of node elevations, meters \\
		elev\_std                     & decimal & Standard deviation of node elevations, meters \\
		grade\_mean                   & decimal & Mean absolute street grade (incline) \\
		grade\_median                 & decimal & Median absolute street grade (incline) \\
		intersect\_count              & integer & Count of physical street intersections \\
		intersect\_count\_clean       & integer & Count of physical street intersections (after merging nodes within 10 meters geometrically) \\
		intersect\_count\_clean\_topo & integer & Count of physical street intersections (after merging nodes within 10 meters topologically) \\
		k\_avg                        & decimal & Avg node degree (undirected) \\
		length\_mean                  & decimal & Mean street segment length (undirected edges), meters \\
		length\_median                & decimal & Median street segment length (undirected edges), meters \\
		length\_total                 & decimal & Total street length (undirected edges), meters \\
		node\_count                   & integer & Count of nodes \\
		orientation\_entropy          & decimal & Entropy of street bearings \\
		orientation\_order            & decimal & Orientation order of street bearings \\
		pagerank\_max                 & decimal & Maximum PageRank value of any node \\
		prop\_4way                    & decimal & Proportion of nodes that represent 4-way street intersections \\
		prop\_3way                    & decimal & Proportion of nodes that represent 3-way street intersections \\
		prop\_deadend                 & decimal & Proportion of nodes that represent dead-ends \\
		self\_loop\_proportion        & decimal & Proportion of edges that are self-loops \\
		straightness                  & decimal & The inverse of circuity \\
		street\_segment\_count        & integer & Count of street segments (undirected edges) \\
		uc\_names                     & string  & List of city names within this urban area (UCD) \\
		world\_region                 & string  & Major geographical region (UCD) \\
		world\_subregion              & string  & Minor geographical region (UCD) \\
		resident\_pop                 & integer & Total resident population, 2015 (UCD) \\
		area                          & decimal & Area within boundary polygon, km\textsuperscript{2} (UCD) \\
		built\_up\_area               & decimal & Built-up surface area in 2015, km\textsuperscript{2} (UCD) \\
		\bottomrule
	\end{tabular}
\end{table}

\subsection{Indicator Calculation}

Once the models are all assembled, we load each's saved GraphML file with OSMnx to calculate each indicator described in Table \ref{tab:indicators}. These indicators are merged with a set of essential indicators from the UCD and saved as a CSV-formatted file.

Details and descriptions are in order for the interpretability of some fields in this indicators dataset. The \texttt{country} field contains the name of the country in which the urban area wholly or primarily (in the case of transnational urban areas) lies, while \texttt{country\_iso} contains its ISO 3166-1 alpha-3 code for unambiguous identification. The \texttt{core\_city} field contains the name of the urban area's core (typically largest) city and the \texttt{uc\_id} field contains the unique identifier of this urban area in the UCD, allowing downstream users to join these indicators with all of those in the UCD. The \texttt{uc\_names} field contains a list of city names within this urban area, per the UCD. The \texttt{world\_region} and \texttt{world\_subregion} fields contain the urban area's major and minor geographical region, per the UCD. The \texttt{resident\_pop} field contains the UCD's estimated 2015 resident population in the urban area. The \texttt{area} and \texttt{built\_up\_area} fields contain the UCD's boundary polygon area and built-up surface area (both in km\textsuperscript{2}) respectively.

The \texttt{circuity} indicator is the graph's ratio of street lengths to straight-line distances between adjacent nodes, and \texttt{straightness} is its inverse. The former measures how circuitous the street network is on average, whereas the latter measures how closely its streets approximate straight lines \citep{boeing_off_2021}. The \texttt{elev\_mean}, \texttt{elev\_median}, \texttt{elev\_std}, \texttt{elev\_iqr}, and \texttt{elev\_range} represent the calculated mean, median, standard deviation, interquartile range, and range of node elevations, in meters. These provide indicators of the topography underlying the network. The \texttt{grade\_mean} and \texttt{grade\_median} fields represent the calculated mean and median street grade absolute values.

The \texttt{intersect\_count} indicator represents the number of street intersections in the urban area---that is, the number of nodes with more than two incident edges in an undirected representation of the graph. The \texttt{intersect\_count\_clean} indicator is calculated by merging intersections within 10 meter buffers of each other geometrically (i.e., 10 meter Euclidean radii) before counting them. This prevents the over-counting of complex intersections. For example, the intersection of two divided roads---each comprising two centerline one-way geometries---creates four nodes and thus would otherwise be counted as four intersections. Roundabouts similarly create multiple intersection points unless consolidated. The 10 meter parameterization has a track record in the literature \citep[e.g.][]{barrington-leigh_global_2020}. The buffer could be customized for each study site to reflect local urban design standards, but for the sake of consistent interpretation across the dataset we use a universal parameterization that works relatively well across all study sites.

The \texttt{intersect\_count\_clean\_topo} indicator is calculated by merging intersections within 10 meters of each other topologically along the network. This prevents topologically remote but spatially proximate nodes from being merged. For example, a street intersection may lie directly below a freeway overpass's intersection with an on-ramp. We would not want to merge these and count them as a single intersection, even though their planar Euclidean distance is approximately zero: in reality, they are distinct junctions in the three-dimensional system of roads. Similarly, in a residential neighborhood, a bollarded street may create a dead-end immediately next to an intersection or traffic circle. We would not want to merge this dead-end with the intersection and connect their edges---they are not adjacent nodes in the graph's topology. These examples illustrate (two-dimensional) geometric proximity, but topological remoteness. Accordingly, in some situations we may expect higher intersection counts in \texttt{intersect\_count\_clean} than \texttt{intersect\_count\_clean\_topo}.

Clustering coefficients measure the extent to which a node's neighbors form a complete graph \citep{jiang_topological_2004,opsahl_clustering_2009}. The \texttt{cc\_avg\_dir} and \texttt{cc\_avg\_undir} indicators are the urban area's directed and undirected unweighted average clustering coefficient. The \texttt{cc\_wt\_avg\_dir} and \texttt{cc\_wt\_avg\_undir} indicators are its directed and undirected length-weighted average clustering coefficient. The \texttt{pagerank\_max} indicator is the maximum PageRank value of any node in the urban area: PageRank ranks nodes' importance based on the structure of their links \citep{agryzkov_algorithm_2012,boeing_multi-scale_2020}. The \texttt{self\_loop\_proportion} measures the urban area's proportion of physical street segments that self-loop.

The \texttt{k\_avg} indicator represents the average node degree of the undirected representation of the graph---that is, on average, how many physical streets (rather than directed edges) are incident to each node. The \texttt{length\_mean} and \texttt{length\_median} indicators are the calculated mean and median physical street segment (i.e., undirected edge) lengths in meters, representing the average and typical linear block lengths. The \texttt{street\_segment\_count} and \texttt{node\_count} fields contain the counts of physical street segments and nodes respectively. The \texttt{prop\_4way}, \texttt{prop\_3way}, and \texttt{prop\_deadend} fields contain the proportions of nodes in the graph that represent four-way intersections, three-way intersections, and culs-de-sac respectively. The \texttt{orientation\_entropy} and \texttt{orientation\_order} indicators represent the calculated entropy of street bearings and their linearized and normalized order, as developed in \citet{boeing_urban_2019}.

\section{Results and Discussion}

\subsection{Open Data Repositories}

All of the resulting street network models, indicators, and metadata have been made publicly available on the Harvard Dataverse, organized within a top-level dataverse\endnote{Dataverse: \url{https://dataverse.harvard.edu/dataverse/global-urban-street-networks/}} collectively comprising approximately 80 gigabytes of data. All of these datasets are freely available for reuse: see the conclusion section for usage notes. All model files (in GeoPackage, GraphML, and node/edge list format) are compressed and zipped at the country level. The top-level dataverse contains five constituent datasets to organize the street network models, indicators, and metadata for retrieval:

\begin{itemize}
	\item Global Urban Street Networks Indicators\endnote{Global Urban Street Networks Indicators repository v2: \url{https://doi.org/10.7910/DVN/ZTFPTB}}: contains the calculated indicators plus essential fields carried over from the UCD for use in downstream analyses. See Table \ref{tab:indicators}.
	\item Global Urban Street Networks Metadata\endnote{Global Urban Street Networks Metadata repository v2: \url{https://doi.org/10.7910/DVN/WMPPF9}}: contains metadata describing the indicators and the node/edge attributes in the model files.
	\item Global Urban Street Networks GraphML\endnote{Global Urban Street Networks GraphML repository v2: \url{https://doi.org/10.7910/DVN/KA5HJ3}}: contains all the GraphML street network model files, compressed and zipped at the country level.
	\item Global Urban Street Networks GeoPackages\endnote{Global Urban Street Networks GeoPackages repository v2: \url{https://doi.org/10.7910/DVN/E5TPDQ}}: contains all the street network model GeoPackage files, compressed and zipped at the country level.
	\item Global Urban Street Networks Node/Edge Lists\endnote{Global Urban Street Networks Node/Edge Lists repository v2: \url{https://doi.org/10.7910/DVN/DC7U0A}}: contains all the street network model node and edge lists in CSV file format, compressed and zipped at the country level.
\end{itemize}

\subsection{Indicator Analysis}

Table \ref{tab:results} presents the results of this workflow by aggregating and summarizing a subset of street network form indicators of particular interest to transportation researchers. It reports population-weighted mean values across all urban areas, aggregated and summarized at the level of the world subregion.

Each table column represents an indicator, some of which are transformed for presentation as follows. The \texttt{Circuity Pct} column subtracts 1 from the \texttt{circuity} indicator and expresses the result as a percent. It thus represents how much more circuitous the streets are than if they were all straight lines. The \texttt{Avg Node Degree}, \texttt{Orientation Order}, and \texttt{Median Street Length} columns present the \texttt{k\_avg}, \texttt{orientation\_order}, and \texttt{length\_median} indicator values respectively. The \texttt{Avg Grade} column expresses the \texttt{grade\_mean} indicator as a percent. The \texttt{Intersect Density} column divides the \texttt{intersect\_count\_clean\_topo} indicator by the \texttt{built\_up\_area} indicator, and thus represents street intersections per km\textsuperscript{2}.

Finally, each of these indicators is aggregated at the subregion-level using the population-weighted mean. Such weighting reflects the average person's exposure by measuring the average indicator value as experienced by the subregion's residents within their urban areas. These population-weighted means are calculated as shown in Equation \ref{eq:pop_wt_mean} where $d_s$ is the population-weighted mean value of indicator $d$ in subregion $s$, $n$ is the number of urban areas in subregion $s$, $i$ indexes those urban areas, $d_i$ is the value of indicator $d$ in urban area $i$, $p_i$ is the population of urban area $i$, and $p_s$ is the total population of subregion $s$.

\begin{equation}
\label{eq:pop_wt_mean}
d_s = \sum_{i=1}^n \frac{p_i}{p_s} d_i
\end{equation}

\begin{landscape}
\begin{table}
\centering
\footnotesize
\caption{Aggregate analysis of urban area street network indicators worldwide: population-weighted mean values, calculated and transformed as described in the text.}
\label{tab:results}
\begin{tabular}{lrrrrrr}
	\toprule
	World Subregion & Circuity Pct & Avg Node Degree & Orientation Order & Median Street Length & Avg Grade Pct & Intersect Density \\
	\midrule
	Eastern Africa        & 5.28 & 2.80 & 0.12 &  82.87 & 3.19 & 266.88 \\
	Middle Africa         & 3.81 & 2.89 & 0.12 &  93.97 & 3.16 & 171.56 \\
	Northern Africa       & 3.66 & 2.90 & 0.19 &  49.76 & 2.56 & 284.10 \\
	Southern Africa       & 8.59 & 2.82 & 0.05 &  77.72 & 3.09 & 151.13 \\
	Western Africa        & 4.30 & 2.82 & 0.13 &  92.75 & 2.48 & 159.98 \\
	Eastern Asia          & 4.86 & 2.96 & 0.22 & 168.96 & 2.14 &  45.74 \\
	South-Central Asia    & 6.10 & 2.65 & 0.20 & 138.83 & 1.97 & 133.93 \\
	South-Eastern Asia    & 6.22 & 2.60 & 0.14 &  64.06 & 2.68 & 190.12 \\
	Western Asia          & 4.97 & 2.95 & 0.13 &  70.04 & 3.46 & 197.08 \\
	Eastern Europe        & 4.97 & 2.84 & 0.09 & 114.48 & 2.23 &  39.04 \\
	Northern Europe       & 6.50 & 2.47 & 0.02 &  65.94 & 2.94 &  97.23 \\
	Southern Europe       & 5.93 & 2.86 & 0.05 &  62.99 & 4.09 & 115.94 \\
	Western Europe        & 6.40 & 2.79 & 0.03 &  78.11 & 2.60 &  72.55 \\
	Caribbean             & 5.62 & 2.82 & 0.10 &  76.38 & 3.14 & 118.38 \\
	Central America       & 3.89 & 2.88 & 0.17 &  62.80 & 4.06 & 166.02 \\
	South America         & 3.68 & 3.03 & 0.15 &  73.22 & 3.85 & 197.76 \\
	Northern America      & 6.71 & 2.87 & 0.32 &  99.95 & 2.50 &  53.23 \\
	Australia/New Zealand & 5.92 & 2.73 & 0.19 &  82.33 & 4.47 &  64.84 \\
	Melanesia             & 8.50 & 2.51 & 0.03 & 106.07 & 3.67 & 105.17 \\
	\bottomrule
\end{tabular}
\end{table}
\end{landscape}

Southern Africa and Melanesia have the most circuitous street networks, where the weighted-average urban area has streets 8.6\% and 8.5\% more circuitous than straight lines. In contrast, Northern Africa and South America have the least circuitous, at 3.7\% each. South America and Eastern Asia have the highest weighted-average average node degree (3.0 each), a measure of network connectedness, whereas Northern Europe and Melanesia (2.5 each) have the lowest. Orientation order, a spatial signature of coordinated central planning, is highest in Northern America (0.32) and Eastern Asia (0.22) and lowest in Northern Europe (0.02) and Melanesia (0.03).

Eastern Asia (169 meters) and South-Central Asia (139 meters) have the longest weighted-average median street segment lengths. In contrast, Northern Africa (50 meters) and Central America (63 meters) have the shortest. The weighted-average urban areas in Australia/New Zealand (4.5\%) and Southern Europe (4.1\%) have the highest average street grades, indicating cities built on hillier terrain, whereas Eastern (2.1\%) and South-Central Asia (2.0\%) have the lowest, indicating cities built on flatter land. Finally, intersection densities (topologically cleaned) are highest in Northern Africa (284/km\textsuperscript{2}) and Eastern Africa (267/km\textsuperscript{2}) and lowest in Eastern Europe (39/km\textsuperscript{2}) and Eastern Asia (46/km\textsuperscript{2}).

\begin{figure*}[tbh]
	\centering
	\includegraphics[width=0.95\textwidth]{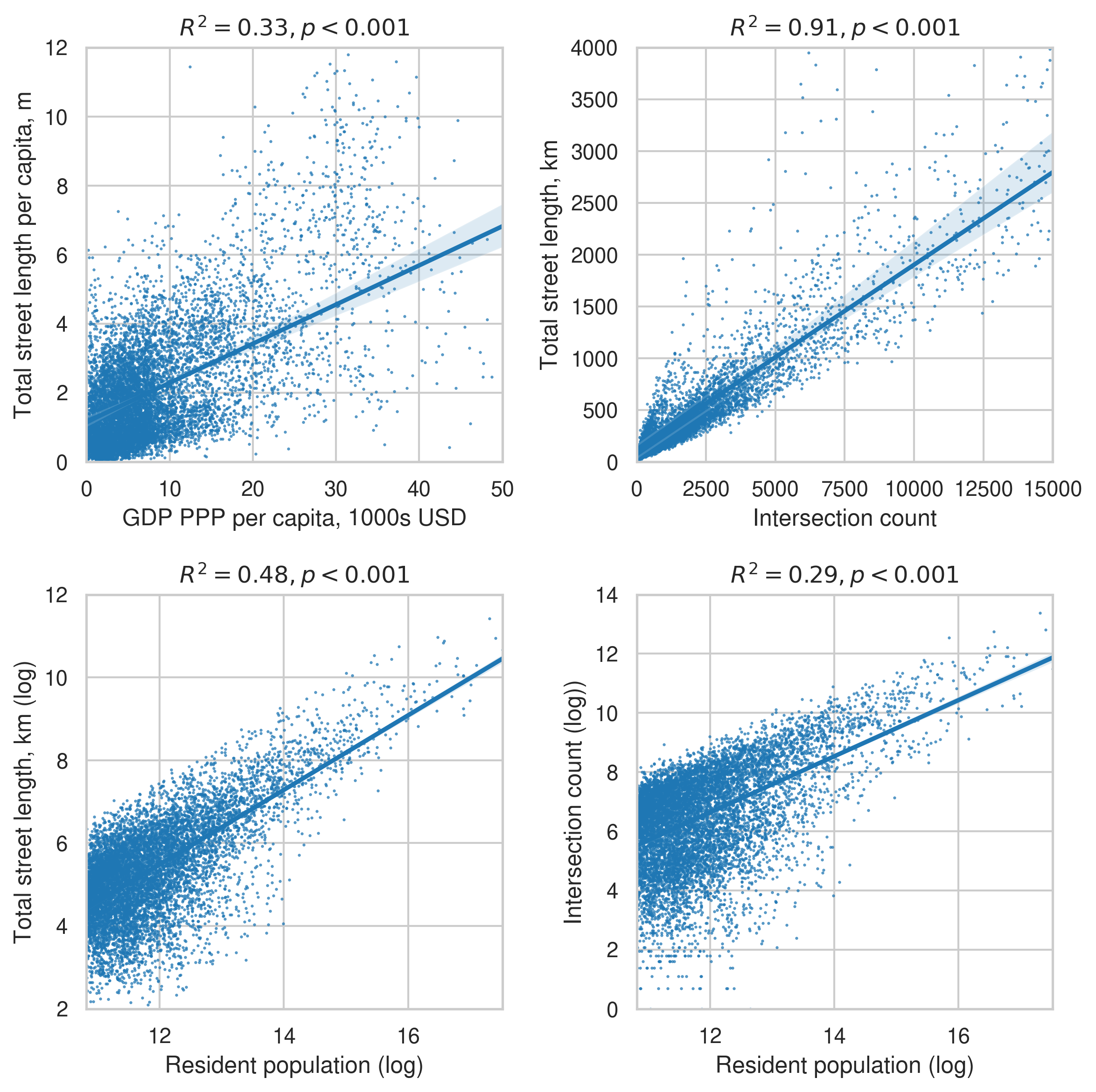}
	\caption{Scatter plots of relationships across all worldwide urban areas with bivariate regression lines and shaded 95\% confidence intervals. Axes constrained to not display all outliers.}
	\label{fig:scatterplots}
\end{figure*}

Figure \ref{fig:scatterplots} visualizes a set of fundamental bivariate relationships (estimated via ordinary least squares with variables transformed as needed for best linear fit) across all urban areas worldwide. Urban areas' intersection counts (topologically cleaned) exhibit a strong linear relationship with their total street lengths (\textit{R}\textsuperscript{2} = 0.91, \textit{p} < 0.001), as expected from theory. A 1 intersection increase in an urban area's intersection count is associated with a 178.6 meter increase in its total street length, ±1.2 meters (margin of error at 95\% confidence).

Total street length and intersection count both scale slightly sublinearly with urban area population, as seen in the log-log plots at the bottom of Figure \ref{fig:scatterplots}. A 1\% increase in urban area population is associated with a 0.90\% (±0.02\%) increase in total street length and a 0.95\% (±0.03\%) increase in intersection count, which makes theoretical sense as residents can share public infrastructure. Similar results have been found in prior studies of a single country \citep[e.g.,][]{bettencourt_origins_2013}.

Urban areas' per capita GDP estimates (i.e., the UCD's 2015 urban area GDP, based on purchasing power parity, in 2011 USD) exhibit a moderate linear relationship with total street length per capita. Across all urban areas, the mean per capita street length is 2.13 meters, and a \$10,000 USD increase in per capita GDP is associated with a 1.13 (±0.03) meter increase in per capita street length. Per capita street length is a common indicator of a city's \enquote{infrastructure accessibility} and this finding provides new evidence consilient with prior smaller-sample findings in the literature: cities with greater wealth and economic activity tend to have more road infrastructure \citep[e.g.,][]{dingil_transport_2018}.

\subsection{Validation}

To be useful in urban science and practice, these models must faithfully represent the real world. Validation needs to be considered from three perspectives.

First, the source data themselves must accurately represent the real-world. In terms of study site demarcation, the UCD derives from the state-of-the-art GHSL developed in tandem by international authorities for this purpose. Numerous researchers have investigated OpenStreetMap's urban street accuracy and completeness \citep{barron_comprehensive_2014,basiri_quality_2016,corcoran_analysing_2013,girres_quality_2010,haklay_how_2010,neis_street_2011,zielstra_assessing_2013}. These data are not perfect. \citet{barrington-leigh_worlds_2017} note that OpenStreetMap was particularly incomplete in China as of 2016 due to restrictions on private surveying and publishing geospatial data. The Chinese models and indicators may in turn suffer from missing source data, and this could partly explain the low intersection density in Eastern Asia in Table \ref{tab:results} as China accounts for 79\% of the subregion's urban population. Nevertheless, OpenStreetMap is the current state-of-the-art for international street network analysis and represents the best available global data today.

The second validation perspective considers how well the resulting graph models represent the OpenStreetMap street network. Are the models constructed properly? Does the workflow introduce errors? The open source modeling software, OSMnx, has been downloaded and installed over 300,000 times from the Anaconda package repository, generating a large test bed of users continuously vetting its functionality. To further assess these models, this study tests the resulting repository's data quality against the original OpenStreetMap source data as a reference dataset by adapting the methodologies of \citet{barrington-leigh_worlds_2017} and \citet{karduni_protocol_2016}. It randomly samples 100 US urban areas and 100 non-US urban areas then manually compares each with the OpenStreetMap source data. The results conform to expectations given the source data and OSMnx's parameterization. Every graph model is also tested to ensure it can be loaded, analyzed, and routed, confirming that this study's computational workflow created a functioning model as described in the methods section.

Third, the elevation data are validated by comparing the ASTER, SRTM, and Google values for each node. Across all 37 million nodes worldwide, the elevation values between these three sources exhibit high correlation (all \textit{r} > 0.999, \textit{p} < 0.001). Table \ref{tab:elevation_validation} summarizes their pairwise differences. The ASTER and SRTM node elevation values differ by only 19 centimeters on average, and the median node has the same elevation value across both. The ASTER and SRTM differences with the Google validation values are greater but both the mean and median remain near or under 1 meter. The high min/max values represent noisy outliers that our node elevation selection process mitigated: across all urban areas, the median node's \textit{selected} elevation value differs from Google by only 32 centimeters. Finally, we compare our street networks' \texttt{elev\_mean} indicator with the UCD's estimated average elevation for each urban area. They strongly correlate (\textit{r} > 0.999, \textit{p} < 0.001) and the median difference between the two is 16 centimeters. Overall, the high correlations and small pairwise differences demonstrate the tight correspondence between the different elevation values and lend confidence to using the ASTER and SRTM open data in these models.

\begin{table}[tbp]
	\centering
	\footnotesize
	\caption{Summary statistics of pairwise differences in elevation (meters) from ASTER, SRTM, and Google across all nodes worldwide.}
	\label{tab:elevation_validation}
	\begin{tabular}{lrrr}
		\toprule
		{}      & ASTER$-$SRTM & ASTER$-$Google & SRTM$-$Google \\
		\midrule
		Mean    &        -0.19 &           0.99 &          1.18 \\
		Std Dev &         7.83 &           7.90 &          3.26 \\
		Min     &      -319.00 &        -319.01 &       -276.55 \\
		25\%    &        -5.00 &          -3.55 &         -0.52 \\
		50\%    &         0.00 &           1.00 &          0.57 \\
		75\%    &         4.00 &           5.33 &          2.57 \\
		Max     &       441.00 &         442.34 &        138.83 \\
		\bottomrule
	\end{tabular}
\end{table}

\section{Conclusion}

\subsection{New Models and Indicators for Street Network Science}

This article presented new methods to model and analyze the street network of each urban area in the world. It used open-source tools and open data to build these models and calculate geometric and topological indicators of street network form. All of its Python source code and resulting data have been deposited in open repositories for public reuse. This represents the first such comprehensive repository of ready-to-use urban area street network models and indicators worldwide.

Analyzing these indicators reveals a snapshot of street network form around the world. We find that both total street length and intersection count scale sublinearly with urban area population. Higher per capita GDP in an urban area is associated with higher per capita total street length. We also find that elevation open data from ASTER and SRTM compare well to Google's closed source, commercial data. All three sources' elevation values correlate strongly. The median pairwise difference between the nodes' elevation values and the Google validation values is only 32 centimeters, and the median pairwise difference between the urban areas' mean street network elevations and the UCD mean elevations is only 16 centimeters.

The open data repositories generated by this study fill two needs in the research community. First, the network models allow researchers to quickly engage in graph-theoretic street network analyses worldwide without first spending weeks writing their own ad hoc code for data collection and modeling. Second, the indicators data provide the first comprehensive worldwide set of geometric and topological street network form indicators at the urban area scale. Together, these results help democratize street network science, opening up quantitative analyses to urban planners and policymakers with less-technical backgrounds who otherwise may struggle to develop a complete computational analytics workflow themselves.

While OpenStreetMap itself provides incredibly valuable raw data, this project transforms these data into ready-to-use models and indicators through substantial processing. For example, these topologically simplified graphs provide models that correspond much better to graph theory and transportation geography than raw OpenStreetMap data do, and they are much faster to run graph algorithms on because most such algorithms scale with node count. They also include elevation and grade data, which are very sparse on OpenStreetMap and are too often ignored in street network analytics. Researchers and practitioners can use these models to simulate trips, assess network vulnerability to flooding and sea level rise, or measure accessibility to points of interest.

The indicators data offer a useful basket of variables for cross-sectional, worldwide studies of street network form, and the topologically-consolidated intersection counts and densities contribute a more theoretically-sound measure than traditional node counts or purely geometric consolidation can. Researchers and practitioners can use these indicators to estimate relationships between street network characteristics and transport-sector greenhouse gas emissions around the world, compare per capita road infrastructure provision across urban areas, or scorecard different cities' street network compactness and connectivity for sustainability initiatives.

\subsection{How to Use These Models and Indicators}

Finally, this article concludes with a few notes on reuse. The indicators data can be loaded with any data analysis tool. The street network models can be loaded with most GIS and network analysis tools, including OSMnx.

Each graph model file is named as \texttt{city\_name-uc\_id.extension}, where \texttt{uc\_id} is the urban area's unique identifier in the UCD. The \texttt{uc\_id} field thus links each graph model file to its urban area in the UCD as well as to its row in the indicators dataset. For example, within the \texttt{china-CHN.zip} file from the Global Urban Street Networks GraphML repository are all of China's urban areas' GraphML files, including \texttt{beijing-10687.graphml}. This GraphML file contains the street network model of Beijing and its \texttt{uc\_id} is \texttt{10687}. As such, \texttt{10687} is Beijing's unique identifier in the indicators dataset and in the UCD.

All of the code used for modeling and analysis in this study is open source and available on GitHub\endnote{Modeling and analysis source code: \url{https://github.com/gboeing/street-network-models}} for public inspection, adaptation, and reuse. Comprehensive documentation of OSMnx and its modules and functions used in this study is available online\endnote{OSMnx documentation: \url{https://osmnx.readthedocs.org}} and OSMnx usage examples, tutorials, and demonstrations are available on GitHub\endnote{OSMnx usage examples: \url{https://github.com/gboeing/osmnx-examples}} for users interested in working with this toolkit.

\IfFileExists{\jobname.ent}{\theendnotes}{}

\section*{Acknowledgments}

This work was funded in part by a grant from the Public Good Projects.

\setlength{\bibsep}{0.00cm plus 0.05cm} 
\bibliographystyle{apalike}
\bibliography{GeAn-world-street-networks}
\end{document}